\documentclass{aa}  

\usepackage{graphicx}
\usepackage{txfonts}
\begin{document}

   \title{A global correlation linking young stars, clouds, and galaxies}
   \subtitle{Towards a unified view of star formation}

   \author{I. Mendigut\'\i{}a\inst{1}
   \and
   C.J. Lada\inst{2}
   \and
   R.D. Oudmaijer\inst{3}                
   }          

   \institute{$^{1}$Centro de Astrobiolog\'{\i}a (CSIC-INTA), Departamento de Astrof\'{\i}sica, ESA-ESAC Campus, PO Box 78, 28691 Villanueva de la Ca\~nada, Madrid, Spain. \email{imendigutia@cab.inta-csic.es}\\
   $^{2}$Harvard-Smithsonian Center for Astrophysics, 60 Garden Street, Cambridge, MA 02138, USA.\\
   $^{3}$School of Physics and Astronomy, University of Leeds, Woodhouse Lane, Leeds LS2 9JT, UK.\\
              }

   \date{Received April 5, 2018; accepted July 30, 2018}

% \abstract{}{}{}{}{} 
% 5 {} token are mandatory
 
  \abstract
   {The star formation rate (SFR) linearly correlates with the amount of dense gas mass (M$_{dg}$) involved in the formation of stars both for distant galaxies and clouds in our Galaxy. Similarly, the mass accretion rate ($\dot{M}_{\rm acc}$) and the disk mass (M$_{disk}$) of young, Class II stars are also linearly correlated.}
   {We aim to explore the conditions under which the previous relations could be unified.}
   {Observational values of SFR, M$_{dg}$, $\dot{M}_{\rm acc}$, and M$_{disk}$ for a representative sample of galaxies, star forming clouds, and young stars have been compiled from the literature. Data were plotted together in order to analyze how the rate of gas transformed into stars and the mass of dense gas directly involved in this transformation relate to each other over vastly different physical systems.}
   {A statistically significant correlation is found spanning $\sim$ 16 orders of magnitude in each axis, but with large scatter. This probably represents one of the widest ranges of any empirical correlation known, encompassing galaxies that are several kiloparsec in size, parsec-size star-forming clouds within our Galaxy, down to young, pre-main sequence stars with astronomical unit-size protoplanetary disks. Assuming that this global correlation has an underlying physical reason, we propose a bottom-up hypothesis suggesting that a relation between $\dot{M}_{\rm acc}$ and the total circumstellar mass surrounding Class 0/I sources (M$_{cs}$; disk+envelope) drives the correlation in clouds that host protostars and galaxies that host clouds. This hypothesis is consistent with the fact that the SFRs derived for clouds over a timescale of 2 Myr can be roughly recovered from the sum of instantaneous accretion rates of the protostars embedded within them, implying that galactic SFRs averaged over $\sim$ 10-100 Myr should be constant over this period too. Moreover, the sum of the circumstellar masses directly participating in the formation of the protostellar population in a cloud likely represents a non-negligible fraction of the dense gas mass within the cloud.}
   {If the fraction of gas directly participating in the formation of stars is $\sim$ 1-35$\%$ of the dense gas mass associated with star-forming clouds and galaxies, then the global correlation for all scales has a near unity slope and an intercept consistent with the (proto-)stellar accretion timescale, M$_{cs}$/$\dot{M}_{\rm acc}$. Therefore, an additional critical test of our hypothesis is that the $\dot{M}_{\rm acc}$-M$_{disk}$ correlation for Class II stars should also be observed between $\dot{M}_{\rm acc}$ and M$_{cs}$ for Class 0/I sources with similar slope and intercept.}

\titlerunning{A global correlation linking young stars, clouds, and galaxies}
   \maketitle
%
%-------------------------------------------------------------------

\section{Introduction}
\label{Sect:Intro}
Star formation encompasses a broad range of physical regimes that account for the collapse of gas to form smaller structures: from star-forming galaxies to individual young stars surrounded by accreting envelopes and disks, passing through intermediate scales including giant molecular clouds, clumps, and dense cores. 

Almost 60 years ago, \citet{Schmidt59} proposed that the star formation rate (SFR) is a power-law function of the total gas density (atomic and molecular), $\sum$$_{SFR}$ = $K$$\sum$$_{gas}^N$ \citep{Kennicutt98}, with $\sum$$_{SFR}$ and $\sum$$_{gas}$ the SFR and the total gas mass (M$_{gas}$) per unit surface area, and $K$ and $N$ the appropriate constants. The Schmidt-Kennicutt law has been observationally confirmed many times when spanning different types of galaxies, revealing a nonlinear correlation with typical values 1 $<$ $N$ $<$ 2 and significant scatter. Using the HCN molecule to trace dense gas mass (M$_{dg}$; n(H$_{2}$) $>$ 10$^4$ cm$^{-3}$), \citet{Gao04} found a roughly linear ($N$ $\sim$ 1) correlation between the spatially integrated SFR and M$_{dg}$ for a wide range of galaxies including normal spirals, luminous, and ultra-luminous infrared galaxies (LIRGs and ULIRGs, respectively). \citet{Wu05} then found that the spatially integrated SFR in distant star-forming clouds in our Galaxy also correlates linearly with the amount of HCN-emitting gas \citep[see also][]{Jackson13, Stephens16}, showing that this correlation also connected smoothly with the one previously found for galaxies. On this basis, \citet{Wu05} suggested that the formation of stars in other galaxies may be understood in terms of the basic units of star formation in the Milky Way (MW), which they identified as the clouds massive enough to fully sample the initial mass function (IMF) by hosting massive young stars. The measurements were extended to nearby molecular clouds with and without massive stars using dust extinction to more accurately trace dense gas masses and direct counts of young stars to more robustly measure the SFRs \citep{Lada10,Lada12}. These measurements confirm that the correlation is linear and spreads approximately ten orders of magnitude in each axis. In summary, strong lines of evidence support the linearity of the SFR-M$_{dg}$ correlation and that this connects individual Galactic clouds and the various types of star-forming galaxies \citep[see also the recent works in][]{Shimajiri17,Tan18}.

The previous view somewhat neglects the role of the actual units in star formation, which are the individual, forming stars. This can be partly attributed to the fact that the SFR, defined as the rate at which gas transforms into stars (in units of M$_{\odot}$ yr$^{-1}$), characterizes physical systems that typically include $\sim$ 10-10$^{11}$ objects. However, there is an equivalent parameter for individual stars that also accounts for the gas-to-star transformation by measuring the rate at which gas falls from the circumstellar environment onto the stellar surface: the stellar mass accretion rate ($\dot{M}_{\rm acc}$, in units of M$_{\odot}$ yr$^{-1}$ too). Direct estimates of $\dot{M}_{\rm acc}$ are only available for optically visible pre-main sequence objects. These are young stars in a relatively advanced stage of early stellar evolution \citep[Class II from the classification of][]{Lada87}, when the optically thick envelope has dissipated and direct observations of the remaining circumstellar disks and the stellar accretion shocks are possible. Their gas disk masses (M$_{disk}$; n(H$_{2}$) $>$ 10$^5$ cm$^{-3}$) are in turn inferred either from observations of different molecules or from submillimeter and millimeter dust continuum emission and an assumed gas-to-dust ratio. Different works indicate that so far the latter method provides more accurate measurements of M$_{disk}$ \citep{Manara16,Rosotti17}. In fact, $\dot{M}_{\rm acc}$ and dust-based M$_{disk}$ measurements are also linearly correlated for large samples of optically visible young stars \citep{Najita07,Mendi12,Najita15}, a result that has been unambiguously confirmed from recent Atacama Large Millimeter/submillimeter Array (ALMA) data at least for the star-forming regions Lupus and Cha I, despite the large scatter \citep{Manara16,Mulders17}. Although one could be tempted to interpret the $\dot{M}_{\rm acc}$-M$_{disk}$ correlation of stars only in terms of small-scale physical processes in the circumstellar disks, this correlation resembles a scaled-down version of the above-mentioned linear correlation linking local clouds and other galaxies. 

This paper shows that the linear correlation that links galaxies and star-forming clouds within our Galaxy (through SFR and M$_{dg}$) can indeed be connected down to individual, forming stars when physically equivalent measurements ($\dot{M}_{\rm acc}$ and M$_{disk}$) are used for the latter objects. We propose that such global correlation has an underlying explanation that may ultimately contribute to a unified view of star formation across an enormous range of astronomical scale. Section \ref{Sect:stars-clouds} shows that the correlations for young stars and clouds in the MW can be connected. We discuss the origin of such a single correlation under the hypothesis that the relation observed in clouds is driven by the individual (mainly Class 0/I) stars that form them. Section \ref{Sect:global} extends the previous finding and discussion, including data for other galaxies. Finally, Sect. \ref{Sect:conclusions} summarizes the main conclusions. 

\section{Correlation between stars and clouds}
\label{Sect:stars-clouds}
\begin{figure}
%[!hbtp]
\centering
 \includegraphics[width=8.5cm,clip=true]{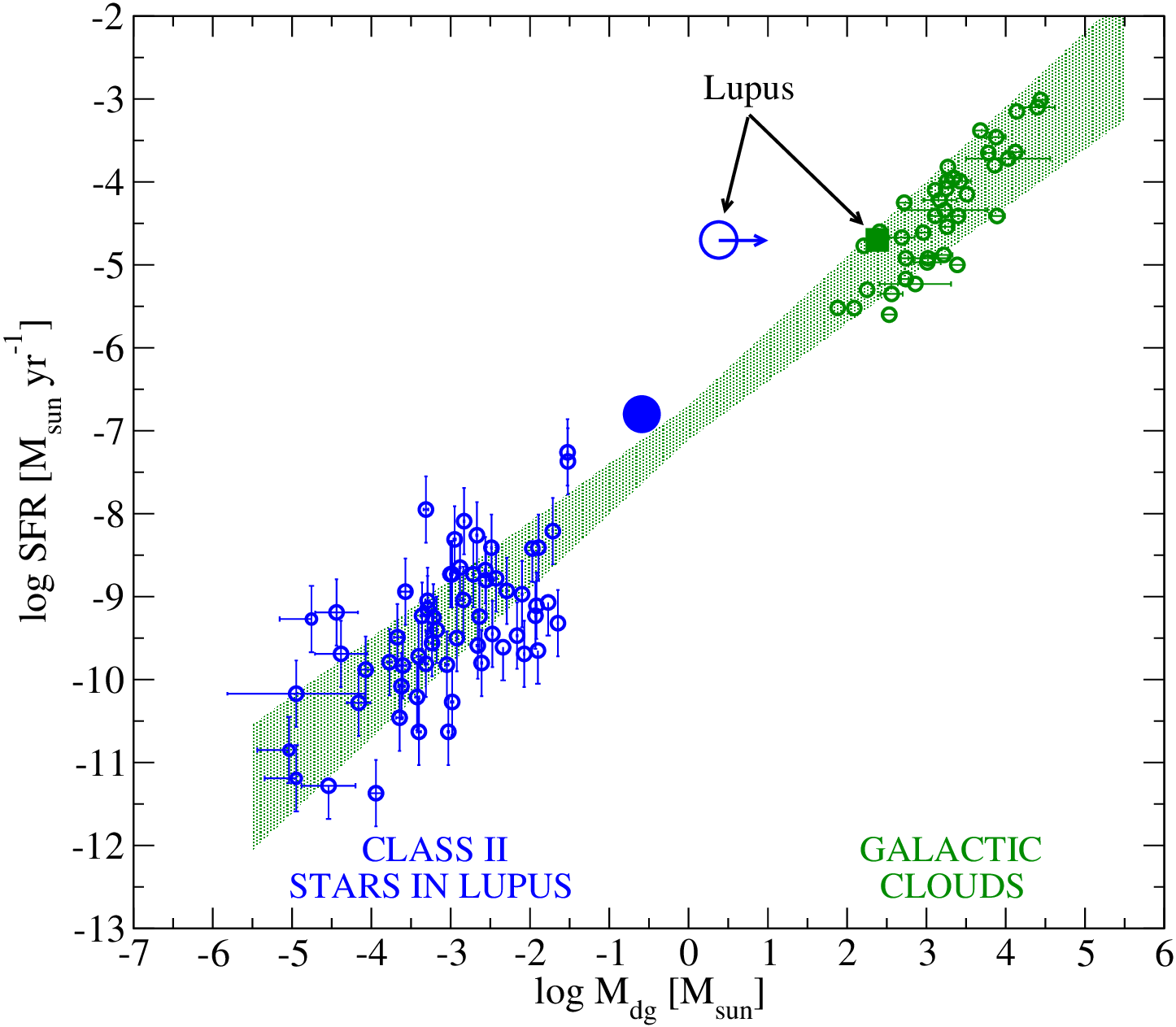}
\caption{Star formation rate versus dense gas mass for distant and nearby Galactic star-forming clouds \citep[green circles with and without horizontal error bars, respectively;][]{Wu10,Lada10} and mass accretion rate versus disk mass in Class II stars in Lupus \citep[blue circles, upper limits as one-sided bars;][]{Manara16}. Error bars are indicated when provided in the original works. The 3$\sigma$ linear fit for stars and clouds is indicated with the green dotted region. The green solid square is the direct measurement of SFR and M$_{dg}$ in Lupus. The blue solid circle is the sum of the accretion rates and disk masses of the Class II stars in Lupus, and the blue open circle includes 20 Class 0/I stars in that cloud with an average accretion rate of 1$\times$10$^{-6}$ M$_{\odot}$ yr$^{-1}$ and a minimum circumstellar mass of 0.1M$_{\odot}$, that is, a maximum displacement in the x-axis given by $f$ = 1$\%$ of the total dense gas mass in Lupus (see text).}
\label{Fig:stars-clouds}
\end{figure}

Figure \ref{Fig:stars-clouds} shows the SFR versus M$_{dg}$ for 39 distant (a few $Kpc$) and nearby (a few hundred $pc$) star-forming clouds in the Galaxy. Only the 28 distant clouds with log M$_{dg}$ $>$ 2.50 (log L$_{HCN}$ $>$ 1.50) from the original sample in \citet{Wu10} have been considered (see Appendices \ref{Sect:appendixa} and \ref{Sect:appendixb}), and most contain HII regions and high-mass young stars. The nearby Galactic clouds are from \citet{Lada10}. The bottom left region shows the $\dot{M}_{\rm acc}$ and M$_{disk}$ values for the nearly complete sample of 62 Class II, optically visible, low-mass young stars in the Lupus I and III star-forming clouds ($\sim$ 150 $pc$) from \citet{Manara16}. The axes are labeled in terms of SFR and M$_{dg}$ for all sources, as often used in studies of star formation scaling laws. The original conversion factors between the observations and the physical parameters compared have been kept and the data considered are representative of each regime, which is enough for our purposes. Appendix \ref{Sect:appendixa} describes how the parameters were derived, and Appendix \ref{Sect:appendixc} shows data for additional samples.

The overall trend in Fig. \ref{Fig:stars-clouds} is confirmed from the Spearman's probability of false correlation using the partial correlation technique \citep[e.g.,][]{Wall03}, which accounts for the fact that the plotted values were computed from a squared-distance factor that depends on the stars and clouds. Its small value ($p$ = 9.5$\times$10$^{-23}$ $<$$<$ 0.05) implies that the correlation is statistically significant and not driven by the d$^2$ factor. Least-squares fitting including 3$\sigma$ errors provides log SFR = -6.9($\pm$0.2) + 0.80($\pm$0.1)log M$_{dg}$, with a correlation coefficient of 0.95. That the slope for the whole sample is not exactly unity is mainly due to the fact that the intercepts for stars and clouds are different. Also relevant is that the y-axis scatter is maximum for stars ($\sim$ $\pm$ 1.5 dex) and decreases for clouds ($\sim$ $\pm$ 0.75 dex). Both topics are discussed in Sects. \ref{Sect:Mdg} and \ref{Sect:global}.

In order to understand the correlation, we propose the hypothesis that the trend in clouds is driven by a relation between the mass accretion rate onto the central object and the total circumstellar gas mass (M$_{cs}$ hereafter) surrounding the individual stars in each cloud. The total circumstellar gas mass refers to M$_{disk}$ for Class II objects, and to M$_{disk}$ plus physically associated material in a surrounding dense core for Class 0/I objects. In the latter case, M$_{disk}$ is usually insignificant compared to M$_{cs}$, but both M$_{cs}$ and M$_{disk}$ refer to the surrounding dense gas mass that is directly participating in the formation of Class 0/I and Class II stars, respectively. To develop this idea in detail we need to consider the extent to which the numbers plotted in Fig. \ref{Fig:stars-clouds} are comparable (see Appendix \ref{Sect:appendixa}). In particular, if our bottom-up hypothesis is correct, the SFR and M$_{dg}$ of a given cloud should scale with the sum of the individual $\dot{M}_{\rm acc}$ and M$_{cs}$ values of all stars in that cloud. 

\subsection{Link between SFR and $\dot{M}_{\rm acc}$}
\label{Sect:SFR}
The $\dot{M}_{\rm acc}$ and SFR estimates in the vertical axis of Fig. \ref{Fig:stars-clouds} are compared in the following. The sum of the accretion rates of all Class II stars in Lupus from \citet{Manara16} is $\sim$ 1.7 $\times$ 10$^{-7}$ M$_{\odot}$ yr$^{-1}$ (log ($\Sigma$$\dot{M}_{\rm acc}$) $\sim$ -6.8), which is more than an order of magnitude smaller than the SFR directly estimated for that cloud \citep[log SFR $\sim$ -4.7;][]{Lada10}. This difference is not only caused by the fact that Class 0/I sources are not included in the sample of \citet{Manara16}, but also because of the different timescales involved in the calculation of SFR and $\dot{M}_{\rm acc}$: the averaged one during the past $\sim$ 2 Myr for the clouds, and the actual (virtually instantaneous) conversion of gas into individual young stars. The amount of gas (per time unit) transformed into a given star during its age (t$_*$) is given by (1/t$_*$) $\times$ $\int_{0}^{t_*}$$\dot{M}_{\rm acc}$(t) dt = (1/t$_*$) $\times$ $\int_{0}^{t_*}$$\dot{M}_{\rm acc}^0$ e$^{-t/\tau_{*}}$ dt = ($\dot{M}_{\rm acc}^0$$\tau$$_*$/t$_*$) $\times$ (1 - e$^{-t_*/\tau_{*}}$), $\dot{M}_{\rm acc}^0$ being the initial mass accretion rate and $\tau$$_*$ the dissipation timescale.\footnote{Three different timescales are used in this work: the ``dissipation timescale'' $\tau$$_*$, as introduced here for Class 0/I and Class II stars; the ``accretion timescale'' $\tau$ = M$_{cs}$/$\dot{M}_{\rm acc}$ (which is M$_{disk}$/$\dot{M}_{\rm acc}$ for Class II stars); and the ``depletion timescale'' M$_{dg}$/SFR, for clouds and galaxies.} Neglecting possible ``FUOr'' and ``EXOr'' events of episodic accretion \citep{Audard14} and focusing on the overall, smooth decrease, an exponential decay with $\tau$$_*$ typically between 1 and 3 Myr is estimated from observations of Class II stars \citep{Fedele10,Mendi12}, and at least one order of magnitude smaller for Class 0/I sources \citep{Bontemps96}. The SFR of a cloud during the past few Myr would then be the sum of the previous expression for all stars in that cloud, which is dominated by the term $\dot{M}_{\rm acc}^0$. 

Direct observational estimates of $\dot{M}_{\rm acc}^0$ are difficult because of the embedded nature of the stars at their initial stages. The empirically based value consistent with most star-forming regions and commonly adopted is 10$^{-4}$ $<$ $\dot{M}_{\rm acc}^0$ (M$_{\odot}$ yr$^{-1}$) $<$ 10$^{-6}$ \citep[e.g.,][]{Palla00,Padoan14}, for which -3 $<$ log SFR $<$ -6. This range is in rough agreement with direct estimates for the clouds. Indeed, within a cloud the actual SFR derived from the sum of the individual $\dot{M}_{\rm acc}$ values is dominated by the high accretion rate of the Class 0/I sources. This is also expected to be of the order of $\dot{M}_{\rm acc}^0$ for individual protostars and, in sum, matches the direct estimates of the SFR. For instance, the typical accretion rate of a Class 0/I star with an IMF stellar mass of $\sim$ 0.3 M$_{\odot}$ and age of $\sim$ 0.3 Myr \citep[e.g.,][]{Padoan14} is given by the ratio between both values, 1$\times$10$^{-6}$ M$_{\odot}$ yr$^{-1}$. Given that Lupus has $\sim$ 20 Class 0/I sources \citep{Merin08,Benedettini12}, the aggregate value $\Sigma$$\dot{M}_{\rm acc}$ is $\sim$ 2 $\times$10$^{-5}$ M$_{\odot}$ yr$^{-1}$ (log($\Sigma$$\dot{M}_{\rm acc}$) $\sim$ -4.7), which coincides with the direct estimate of the total SFR for that region averaged over 2 Myr \citep{Lada10}. Moreover, for three molecular clouds for which low-mass protostars have been reasonably well sampled, we find 329, 139, and 43 Class 0/I sources in Orion A, Perseus, and California \citep{Megeath12,Carney16,Lewis16,Lada17}. This gives $\Sigma$$\dot{M}_{\rm acc}$ = 3.3$\times$10$^{-4}$, 1.4$\times$10$^{-4}$, and 4.3$\times$10$^{-5}$ M$_{\odot}$ yr$^{-1}$, which are within a factor $\leq$ 2 the SFRs of the same clouds estimated considering all young stars and a 2 Myr timescale: 7.1$\times$10$^{-4}$, 1.5$\times$10$^{-4}$ and 7.0$\times$10$^{-5}$ M$_{\odot}$ yr$^{-1}$, respectively \citep{Lada10}. This close agreement again supports the idea that the SFRs directly estimated for entire populations and averaged over 2 Myr are recovered from the sum of the ``instantaneous'' (0.3 Myr) accretion rates when those are dominated by Class 0/I stars, that is, $<$SFR$>$$_{2Myr}$ = $\Sigma$$\dot{M}_{\rm acc}^0$ $\equiv$ SFR$_{0}$. In other words, the SFR is essentially constant over timescales of 2 Myrs, approximately ten times the typical protostellar lifetime. 

\subsection{Link between M$_{dg}$ and M$_{cs}$}
\label{Sect:Mdg}
Next we consider the gas mass estimates on the horizontal axis of Fig. \ref{Fig:stars-clouds}. The sum of all individual disk masses in Lupus from \citet{Manara16} is $\sim$ 0.25 M$_{\odot}$ (log ($\Sigma$${M}_{disk}$) $\sim$ -0.6), which is again orders of magnitude smaller than the M$_{dg}$ value typically estimated for Lupus \citep[log M$_{dg}$ $\sim$ 2.4;][]{Lada10}. However, the sample in \citet{Manara16} refers to evolved Class II stars that are associated with relatively small amounts of gas. For a more complete accounting of gas associated with young stars in a forming cloud, we need to include the contribution from protostellar (Class 0/I) objects to the dense gas budget (similar to the situation with the SFRs discussed above). Being embedded in dense cores and surrounded by infalling material, Class 0/I sources are associated with significantly higher amounts of dense gas. Indeed, protostellar objects are almost exclusively located in regions where the extinction exceeds 0.8 magnitudes at K band \citep[e.g.,][]{Lombardi14,Zari16,Lada17}, which is the defining lower limit used by \citet{Lada10} for their measurements of M$_{dg}$ in nearby clouds. 

Still, circumstellar dense cores likely only account for some fraction of the dense gas in a cloud, $\Sigma$M$_{cs}$ = $f$M$_{dg}$. Because the boundary between a dense core and the surrounding dense gas in which it is embedded is difficult to determine or even define, the value of $f$ is uncertain.  Assuming that the bottom-up hypothesis is correct, we can extrapolate the roughly linear relation for the Lupus Class II sources to match the observed (total) SFR for the Lupus cloud and then predict the value of $\Sigma$M$_{cs}$ $\sim$ SFR $\times$ $\tau$, where $\tau$ is the accretion timescale of the Class II stars, M$_{disk}$/$\dot{M}_{\rm acc}$. For SFR = 2 $\times$ 10$^{-5}$ M$_{\odot}$ yr$^{-1}$ \citep{Lada10} and $\tau$ ranging between $\sim$ 0.15 and 4.25 Myr \citep[][see also Fig. \ref{Fig:timescales} and the discussion in the next section]{Manara16}, we find 3 $<$ $\Sigma$M$_{cs}$/M$_{\odot}$ $<$ 85 (i.e., 0.012 $<$ $\Sigma$M$_{cs}$/M$_{dg}$ $<$ 0.34), which for the $\sim$ 20 Class 0/I sources in the cloud would mean 0.15 $<$ $<$M$_{cs}$$>$/M$_{\odot}$ $<$ 4.2. That is, the total mass of circumstellar gas directly participating in star formation in Lupus appears to represent between 1.2$\%$ and $\sim$ 34$\%$ of the total dense gas reservoir measured in the cloud, and the gas mass around a protostar directly associated with accretion typically contains between 0.15M$_{\odot}$ and 4.2M$_{\odot}$. Direct measurements of some Class 0/I stars in Lupus indicate that a lower limit for $<$M$_{cs}$$>$ is $\sim$ 0.15M$_{\odot}$ \citep{Mowat17}. This and additional Herschel measurements of M$_{cs}$ in Lupus and other star-forming regions \citep{Rygl13,Konyves15} may be suggesting that $f$ is closer to 1$\%$ than to 35$\%$. 

Generally speaking, the gas mass of protostellar and prestellar dense cores typically ranges between 0.1 and 10 M$_{\odot}$ \citep[e.g.,][]{Enoch08}; assuming that 3 $<$ $<$M$_{cs}$$>$/M$_{\odot}$ $<$ 5  and that $f$ $\sim$ N$_{protostars}$$\times$$<$M$_{cs}$$>$/M$_{dg}$, the recent extinction maps of local molecular clouds in \citet{Lombardi14}, \citet{Zari16}, and \citet{Lada17} indicate that 7$\%$ $<$ $f$(OrionA) $<$ 12$\%$, 22$\%$ $<$ $f$(Perseus) $<$ 33$\%$, and  4$\%$ $<$ $f$(California) $<$ 7$\%$. Therefore, a $f$ value ranging between $\sim$ 1\%\ and 35\% is reasonable. Such a range means that the corresponding log ($\Sigma$$\dot{M}_{\rm acc}$)-log($\Sigma$M$_{cs}$) values for the clouds should be shifted between 0.5 dex ($f$ = 35\%) and 2 dex ($f$ = 1\%) to the left from their log SFR-log M$_{dg}$ values in Fig. \ref{Fig:stars-clouds} (the values for the vertical axis would remain practically equal, as previously argued). This change of intercept would not break the correlation and fits even better with the trend shown by stars (see Sect. \ref{Sect:global}). It is noted that a value of $f$ closer to the upper limit of 35\% is obtained if the prestellar cores are considered. Indeed, Herchel surveys indicate that the mass around these sources can dominate over the circumstellar mass around protostars \citep{Konyves15}. In addition, inflow motions have been detected at least in some prestellar cores at large, sub-parsec scales \citep[e.g.,][]{Tafalla98}.    

In summary, the numbers provided should be taken as a first order approach, not only because of the uncertainties involved, but also because it has been assumed that the accretion timescale for Class 0/I sources is similar to that for Class II stars (as discussed below). In any case, our arguments strongly suggest that the sum of individual M$_{cs}$ measurements of Class 0/I sources represents a non-negligible contribution to the total M$_{dg}$ estimates for clouds from extinction maps, in agreement with the bottom-up hypothesis. 

\subsection{Some observational and theoretical implications}
\label{Sect:implications}
Given that the $\dot{M}_{\rm acc}$ and M$_{cs}$ values of Class 0/I sources dominate over more evolved stars, a major consequence is that the observed correlation between $\dot{M}_{\rm acc}$ and M$_{disk}$ in Class II stars could be the tip of the iceberg; our hypothesis implies that the correlation should continue with similar slope and intercept to $\dot{M}_{\rm acc}$ and M$_{cs}$ in earlier Class 0/I stages, filling the gap between the Class II stars and the clouds in Fig. \ref{Fig:stars-clouds}. This means that the accretion timescale for Class 0/I stars, M$_{cs}$/$\dot{M}_{\rm acc}$, should be roughly similar ($\sim$ within $\pm$ 1 dex) to that for Class II objects, M$_{disk}$/$\dot{M}_{\rm acc}$. In other words, the decline of the accretion rate must occur at a pace comparable to that for the dissipation of the circumstellar material in order to keep a similar accretion timescale along the different stages of star formation. The latter has been observed in Class II stars \citep[apart from the Lupus data studied here, see, e.g.,][]{Fedele10}, and has been suggested by observations \citep{Bontemps96} and modeling (see below) for Class 0/I sources independently. In fact, assuming a direct proportionality between accretion and ejection, a linear relation between $\dot{M}_{\rm acc}$ and M$_{cs}$ was inferred based on a comparative study of outflow activity in Class0/I objects \citep{Bontemps96}. 

However, a direct comparison between the Class II and Class 0/I accretion timescales is not feasible so far not only because of the observed scatter in the correlation for the Class II stars in the couple of star-forming regions that have been reliably probed \citep{Manara16,Mulders17}, but especially because such correlation has not been reported for Class 0/I sources, apart from indirect suggestions like the one mentioned above \citep{Bontemps96}. Indeed, accurate $\dot{M}_{\rm acc}$ and M$_{cs}$ values for complete samples of Class 0/I sources are generally lacking. The extrapolation of line-luminosity to accretion-luminosity correlations \citep{Mendi15} to embedded objects are providing promising results \citep{Pomohaci17}. In particular, mid-infrared (IR) emission lines could be useful to provide $\dot{M}_{\rm acc}$ estimates of highly embedded stars \citep{Rigliaco15}, for which the spectroscopic capabilities of the James Webb Space Telescope will be ideally suited. Ongoing submillimeter surveys devoted to characterize the accretion properties of Class 0 sources \citep{Herczeg17} will also be of great value. Regarding M$_{cs}$, ALMA should be able to make an improved census of the protostellar envelope masses in several star-forming regions. 

Similarly, our hypothesis suggests that a relevant theoretical framework must be able to predict a $\dot{M}_{\rm acc}$-M$_{cs}$ correlation at the initial Class 0/I stages. Such type of theories have already been proposed, for example, predicting the $\dot{M}_{\rm acc}$-M$_{disk}$ correlation in Class II stars from the imprint of the parent, turbulent core, either from its rotational properties \citep{Dullemond06} or/and from gravitational instabilities and density perturbations \citep{Vorobyov08}. Alternatively, similar correlations for Class 0/I sources have also been reproduced in models that do not assume ad-hoc initial and boundary conditions for the cores, but emphasize the importance of the turbulent medium considering the cloud as a whole \citep{Padoan14}. According to this view, the observed $\dot{M}_{\rm acc}$-M$_{disk}$ correlation in Class II stars would not be caused by the disk physics but would be evidence of an underlying $\dot{M}_{\rm acc}$-M$_{cs}$ correlation in previous Class 0/I stages. In other words, if our hypothesis is confirmed, viscous disk evolution should be dominant only when the envelopes have dissipated, but its influence on the location of Class II stars in the $\dot{M}_{\rm acc}$-M$_{disk}$ plane -- and thus on the corresponding accretion timescale M$_{disk}$/$\dot{M}_{\rm acc}$ -- would be less relevant as this would have been mainly determined at earlier, embedded stages.    

\section{The global correlation}
\label{Sect:global}
\begin{figure}
%[!hbtp]
\centering
 \includegraphics[width=8.5cm,clip=true]{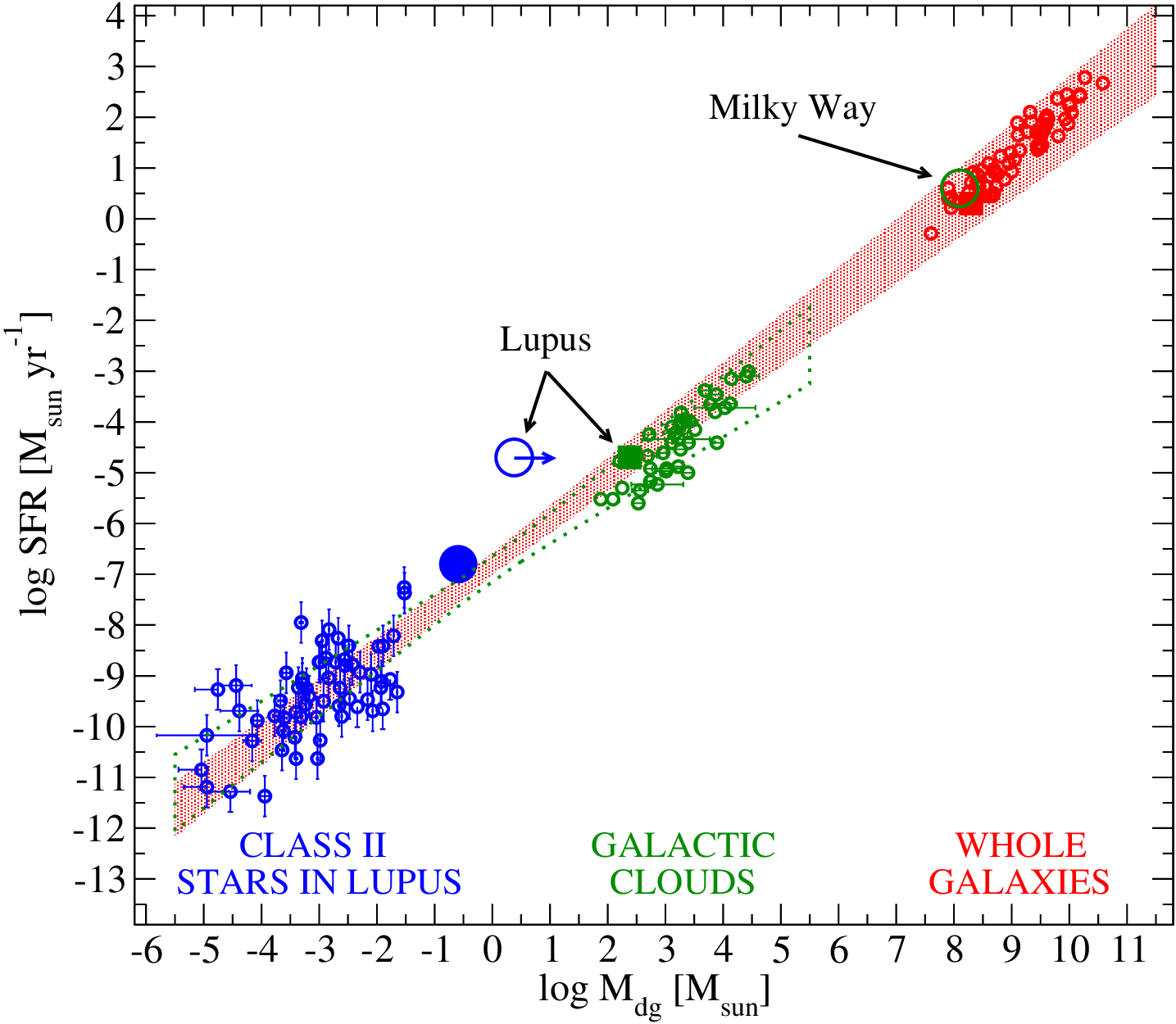}
\caption{Figure \ref{Fig:stars-clouds} including galaxies \citep[red circles;][]{Gao04}. The fit for the stars and clouds is now indicated by the region enclosed within the green dotted line. The red dotted region is the 3$\sigma$ global fit for all sources. The green open circle is the sum of 4$\times$10$^4$ clouds with a typical SFR and M$_{dg}$, and the red solid square is the direct measurement of SFR and M$_{dg}$ in the MW (see text).}
\label{Fig:correlation}
\end{figure}

As introduced in Sect. \ref{Sect:Intro}, strong lines of evidence show that the SFR and M$_{dg}$ values of Galactic clouds and whole galaxies are connected through a linear relation. Figure \ref{Fig:correlation} extends Fig. \ref{Fig:stars-clouds}, adding the SFR and M$_{dg}$ values of the 65 normal spiral galaxies, LIRGS, and ULIRGS at distances between a few and a few hundred Mpc from \citet{Gao04}. The global correlation including all sources overlaps with that for clouds and stars derived in the previous section is also statistically significant ($p$ = 2.6$\times$10$^{-36}$), and has a slope N close to (but still different from) unity (the 3$\sigma$ fit is log SFR = -6.8($\pm$0.2) + 0.88($\pm$0.06)log M$_{dg}$; correlation coefficient = 0.99). As in the previous section, the values for the probability of false correlation and the linear correlation coefficient have been derived from the partial correlation technique \citep[e.g.,][]{Wall03}, indicating that the global correlation is not driven by the wide range of distances to the sources. Therefore, from the statistical perspective the SFR-M$_{dg}$ correlation can be extended from galaxies and clouds to young stars when $\dot{M}_{\rm acc}$ and M$_{disk}$ are used for the latter objects. Once again the inclusion of additional data or the use of different scaling factors between the observations and the physical parameters compared would not have a significant effect on our discussion (Appendices \ref{Sect:appendixa} and \ref{Sect:appendixc}). Moreover, one may still think that any stellar parameter could appear to be similarly correlated on a log-log diagram when such a huge range in both axes is considered, but Appendix \ref{Sect:appendixc} shows that this is not the case.

\cite{Wu05} proposed that the correlation in galaxies can be understood from the sum of basic units within each galaxy, concluding that these are clumps that host high-mass young stars. However, Appendix \ref{Sect:appendixb} shows that distance dependencies and the fact that the correlation extends to local clouds that do not generally form massive stars \citep{Lada10,Lada12} makes this conclusion less robust \citep[see also][]{Stephens16}. Recently, \citet{Shimajiri17} proposed that the relation between kiloparsec-size galaxies and parsec-size clouds may be governed by the “microphysics” of core and star formation along filaments within the clouds. Following these lines of reasoning and based on Fig. \ref{Fig:correlation} and our results in the previous section, we propose that, as basic products of star formation, the (sub-)parsec-size individual protostellar systems may ultimately drive the observed correlation not only for protostellar-hosting clouds but also for cloud-hosting galaxies. 

Our Galaxy follows the global correlation showing log SFR $\sim$ 0.3 \citep{Chomiuk11,Kennicut12} and log M$_{dg}$ $\sim$ 8.3 \citep[e.g.,][and references therein]{Stephens16}. These values can be roughly recovered, for example, by summing up a reasonable value of $\sim$ 4$\times$10$^4$ clouds with a typical log SFR = -4 and log M$_{dg}$ = 3.5. Regardless of the exact numbers, the bottom line is that the SFR and M$_{dg}$ estimates for the MW or any other galaxy could in principle be derived also from the sum of the accreting stellar population within each galaxy, since we have shown in the previous section that the SFR and M$_{dg}$ values for the clouds scale with the sum of the $\dot{M}_{\rm acc}$ and M$_{cs}$ values of the stars hosted by them. For the same reason argued when the SFRs of the clouds and the instantaneous accretion rates of the stars were compared, if our bottom-up hypothesis can be extended to whole galaxies, then their SFRs should be roughly constant along a timescale of 10-100 Myr, given that such a timescale is the one usually considered to estimate galactic SFRs (Appendix \ref{Sect:appendixa}). 

\begin{figure}
%[!hbtp]
\centering
 \includegraphics[width=8.5cm,clip=true]{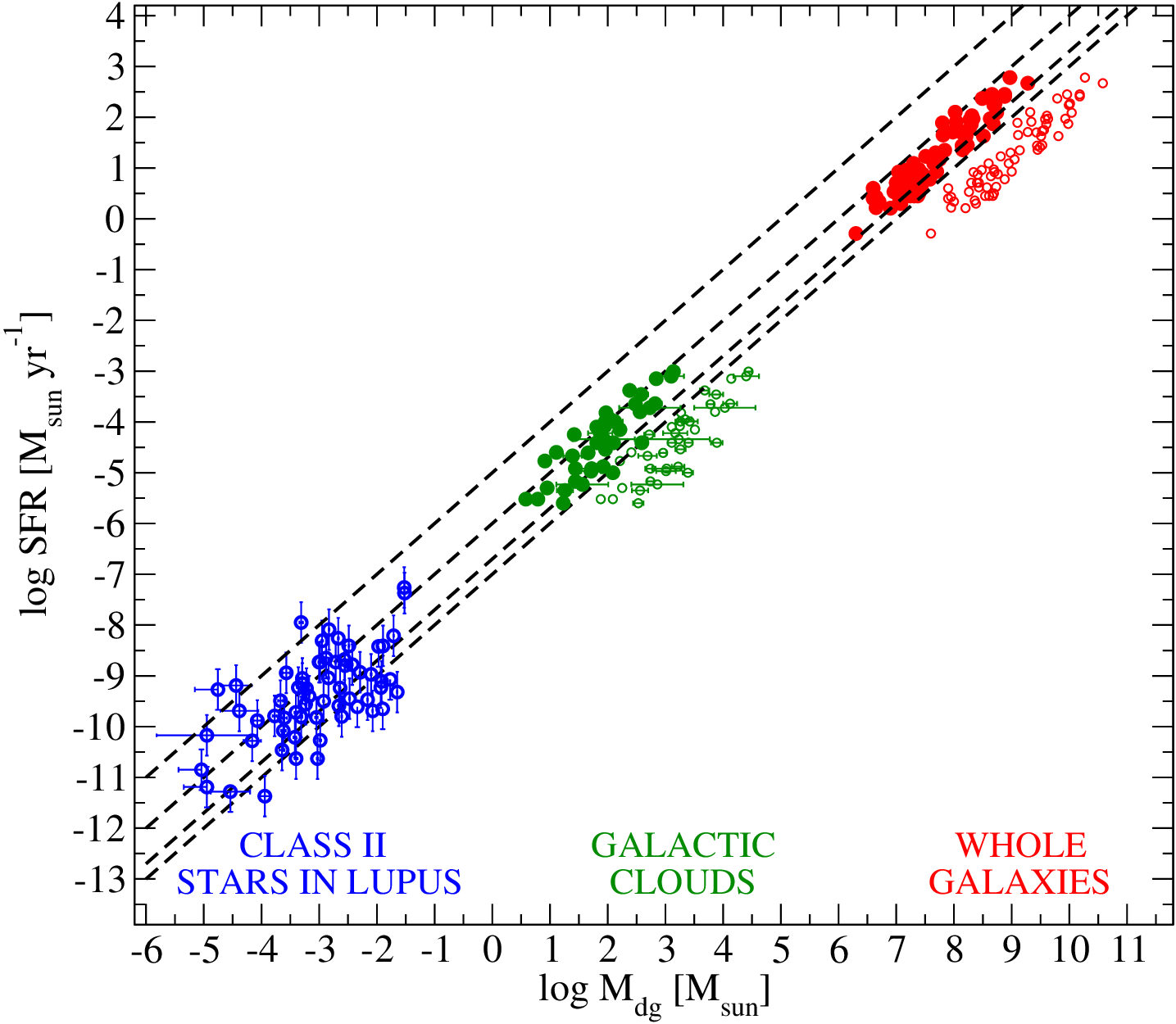}
\caption{Accretion timescales M$_{cs}$/$\dot{M}_{\rm acc}$ = 0.1, 1, 5, and 10 Myr are indicated from top to bottom with the dashed lines. The original data for the clouds and the galaxies in Fig. \ref{Fig:correlation} (green and red open circles) are shifted -1.3 dex in the x-axis (green and red filled circles), corresponding to a fraction of gas directly associated with star formation of 5$\%$ the total dense gas mass measured in those systems.}
\label{Fig:timescales}
\end{figure}

In addition, because estimates for clouds and galaxies are averaged over the entire systems, we show in Appendix \ref{Sect:appendixd} that our bottom-up hypothesis firstly implies that the intercepts in the linear relations derived for stars, clouds, and galaxies individually should be roughly similar, and secondly that the scatter should decrease from stars to galaxies. The second consequence is clearly observed in our data (the y-axis dispersion in galaxies decreases again to $\sim$ $\pm$ 0.5 dex, see also Sect. \ref{Sect:stars-clouds}), in the more complete sample of clouds and galaxies studied by \citet{Stephens16}, or in the very recent work by \citet{Tan18} \citep[but see also][]{Lada12,Shimajiri17}. The first consequence presents some caveats. As discussed in Appendix \ref{Sect:appendixd}, the least-squares intercepts should be similar at least if all stars have comparable $\dot{M}_{\rm acc}$/M$_{cs}$ ratios (see Sect. \ref{Sect:implications}) and if the fraction of dense gas directly involved in star formation is $f$ $\sim$ 100$\%$. Under these assumptions, the gas depletion timescale derived for galaxies and clouds from least-squares fitting (M$_{dg}$/SFR) should be the same as the accretion timescale of the stars (M$_{cs}$/$\dot{M}_{\rm acc}$), but the former should be larger than the latter if $f$ $<$ 100$\%$. Indeed, Fig. \ref{Fig:timescales} shows that whereas an accretion timescale ranging between $\sim$ 0.1 and 10 Myr matches the Class II stars, the original data for the clouds and the galaxies fall displaced to the right, consistent with larger timescales. Shifting the clouds and galaxies between $\sim$ -0.5 and -2 dex in the x-axis provides a better match with the stars, finally resulting in a global correlation with near unity slope and an intercept more consistent with the stellar accretion timescale (e.g., for the shift shown in Fig. \ref{Fig:timescales}, log SFR = -6.3($\pm$0.1) + 0.98($\pm$0.02)log M$_{dg}$). Given that $\Sigma$M$_{cs}$ = $f$M$_{dg}$, the previous shift range corresponds to 0.01 $<$ $f$ $<$ 0.35, suggesting that the amount of gas directly participating in star formation should be roughly between 1$\%$ and 35$\%$ the total dense gas mass measured in clouds and galaxies, in agreement with the observational data discussed in the previous section. Once again the underlying assumption is that the dominating Class 0/I sources have a similar accretion timescale as observed for Class II stars, for which we remark that accurate measurements of $\dot{M}_{\rm acc}$ and M$_{cs}$ for Class 0/I stars are the most critical observations needed to test the bottom-up hypothesis.

\section{Summary and conclusions}
\label{Sect:conclusions}
The gas-to-star transformation rate and the dense gas mass available for star formation show a statistically significant correlation that has been extended from galaxies and Galactic clouds to individual young stars. The correlation spreads $\sim$ 16 orders of magnitude, covering spatial scales from distant (hundreds of Mpc) galaxies with sizes of a few Kpc, to nearby ($\sim$ 150 $pc$) protoplanetary disks with sizes of hundreds of au. Given the difference between the intercepts of the individual linear fits determined for protoplanetary disks, star-forming clouds, and galaxies, it is not clear how physically meaningful the global correlation is. Here we have proposed and explored one possible unifying hypothesis that could explain the global correlation.

Our bottom-up hypothesis is that the correlation between the mass accretion rate and the total circumstellar gas mass surrounding young -- mainly Class 0/I -- stars is the underlying cause of the global correlation on all scales. Two main lines of evidence supporting this scenario have been provided. First, the SFRs of molecular clouds are recovered from the sum of empirically based stellar accretion rates. Second, the sum of the individual circumstellar gas masses appears to represent a relevant fraction of the total dense gas mass within such clouds. If that fraction is between $\sim$ 1\%\ and 35 $\%$ of the dense gas mass in clouds and galaxies, then the global correlation has a near unity slope and intercept consistent with the stellar accretion timescale, M$_{cs}$/$\dot{M}_{\rm acc}$. Indeed, a crucial test of our hypothesis is that the $\dot{M}_{\rm acc}$-M$_{disk}$ linear correlation on astronomical unit scales for Class II stars should also be observed between $\dot{M}_{\rm acc}$ and M$_{cs}$ on (sub-)parsec scales for Class 0/I sources. Our hypothesis also explains that the scatter in the correlation decreases from stars to galaxies, and implies that the average SFR of galaxies remains constant on a timescale of $\sim$ 10-100 Myr. 

The $\dot{M}_{\rm acc}$-M$_{disk}$ and SFR-M$_{dg}$ correlations in stars, clouds, and galaxies have been independently studied from physical contexts relatively isolated from each other. Our result suggests a global approach and that theoretical efforts should consider all scales and physical systems involved in a single correlation. 

\begin{acknowledgements}
The authors thank the anonymous referee for the comments, which have served to improve this manuscript. The authors sincerely acknowledge M. Aberasturi, J.C. Mu\~noz-Mateos, J.M. Mas-Hesse, P. Padoan, N. Cunningham, and B. Montesinos for their help. IM acknowledges the Government of Comunidad Aut\'onoma de Madrid, Spain, which has funded this work through a ``Talento'' Fellowship (2016-T1/TIC-1890). 

\end{acknowledgements}

\begin{appendix} 
\section{Estimates of the different parameters for stars, clouds, and galaxies}
\label{Sect:appendixa} 
This appendix refers to the different ways that the parameters compared in Figs. \ref{Fig:stars-clouds} and \ref{Fig:correlation} were derived.

\textbf{Stars}: Values of $\dot{M}_{\rm acc}$  refer to the instantaneous disk-to-star accretion rates, which were derived by fitting an accretion shock model to the extinction-corrected near-ultraviolet (UV) excess. Values of M$_{disk}$  were calculated from ALMA submillimeter continuum data using typical assumptions of a single dust grain opacity $\kappa$(890 $\mu$m) = 3.37 cm$^2$ g$^{-1}$, a single dust temperature T$_{dust}$ = 20 K, and a gas-to-dust ratio of 100. Further details can be found in \citet{Manara16} and references therein. The typical densities of protoplanetary disks are n(H$_{2}$) $>$ 10$^5$ cm$^{-3}$, which can be as high as $>$ 10$^8$ cm$^{-3}$ in the innermost, densest regions \citep[see, e.g.,][and references therein]{Bergin07}.

\textbf{Nearby clouds}: The SFRs were derived by counting the number of young stars within each cloud (N$_{stars}$), assuming a typical IMF mass (M$_*$ = 0.5M$_{\odot}$), and a common age for all sources (t$_*$ = 2 Myr). Then SFR(M$_{\odot}$ yr$^{-1}$) = N$_{stars}$ $\times$ M$_*$/ t$_*$ =  0.25 $\times$ 10$^{-6}$ N$_{stars}$\footnote{That is, the SFR derived in this way implicitly assumes that the individual stars are actually the basic units driving the correlation, in agreement with the bottom-up hypothesis of this work}. In turn, M$_{dg}$ was calculated from Two Micron All Sky Survey (2MASS) extinction maps of the clouds with A$_K$ $>$ 0.8 magnitudes and a gas-to-dust ratio of 100, which involves gas densities n(H$_{2}$) $>$ 10$^4$ cm$^{-3}$. Further details can be found in \citet{Lada10}. 

\textbf{Distant clouds}: SFR and M$_{dg}$ values were derived from the IR luminosities (L$_{IR}$, from 8 to 1000 $\mu$m) and the HCN (1--0) luminosities (L$_{HCN}$) listed in \citet{Wu10}, following the same procedure as for the galaxies in \citet{Gao04}, described in the following.

\textbf{Galaxies}: SFRs were derived from unresolved measurements of L$_{IR}$, assuming that this is produced primarily from dust heating by O, B, and A stars, a typical Salpeter IMF, and applying SFR (M$_{\odot}$ yr$^{-1}$) = 2 $\times$ 10$^{-10}$ $\times$(L$_{IR}$/L$_{\odot}$). This methodology implicitly assumes that the SFR is constant along a timescale of $\sim$ 100 Myr. Values of M$_{dg}$  were estimated from unresolved measurements of L$_{HCN}$ by applying M$_{dg}$ = $\alpha$$_{HCN}$ $\times$ L$_{HCN}$, with $\alpha$$_{HCN}$ = 10 M$_{\odot}$ (K km s$^{-1}$ pc$^2$)$^{-1}$ a constant derived assuming that L$_{HCN}$ is emitted by the gravitationally bound cloud cores with a brightness temperature of 35 K. The gas densities involved are also n(H$_{2}$) $>$ 10$^4$ cm$^{-3}$. Further details can be found in \citet{Gao04}. It is noted that recent works claim that HCN (1--0) traces densities an order of magnitude smaller \citep{Pety07,Kauffmann17b}. However, \citet{Kauffmann17b} show that even in this case the M$_{dg}$ values derived from L$_{HCN}$ would differ from those in \citet{Gao04} by less than a factor of two, which would not have a significant effect on our results.  

Despite the fact that there are differences in the methodologies, spatial, and temporal scales involved to derive the parameters, all are referred to the same two physical concepts: the rate at which gas is transformed into stars, and the mass of dense gas related to stellar formation. Thus, the comparison between them can and should be carried out. \citet{Lada12} compared the SFRs and  M$_{dg}$ values in clouds and galaxies, as estimated from the different methodologies summarized above \citep[see also, e.g.,][]{Shimajiri17}. Therefore, in our work we have mainly focused on the comparison between $\dot{M}_{\rm acc}$-SFR and M$_{disk}$-M$_{dg}$ in stars and clouds, when the latter were derived from direct counts of stars and extinction maps, respectively.

\section{Re-analysis of the \citet{Wu05} sample of distant Galactic clouds}
\label{Sect:appendixb}
\begin{figure}
%[!hbtp]
\centering
 \includegraphics[width=8.5cm,clip=true]{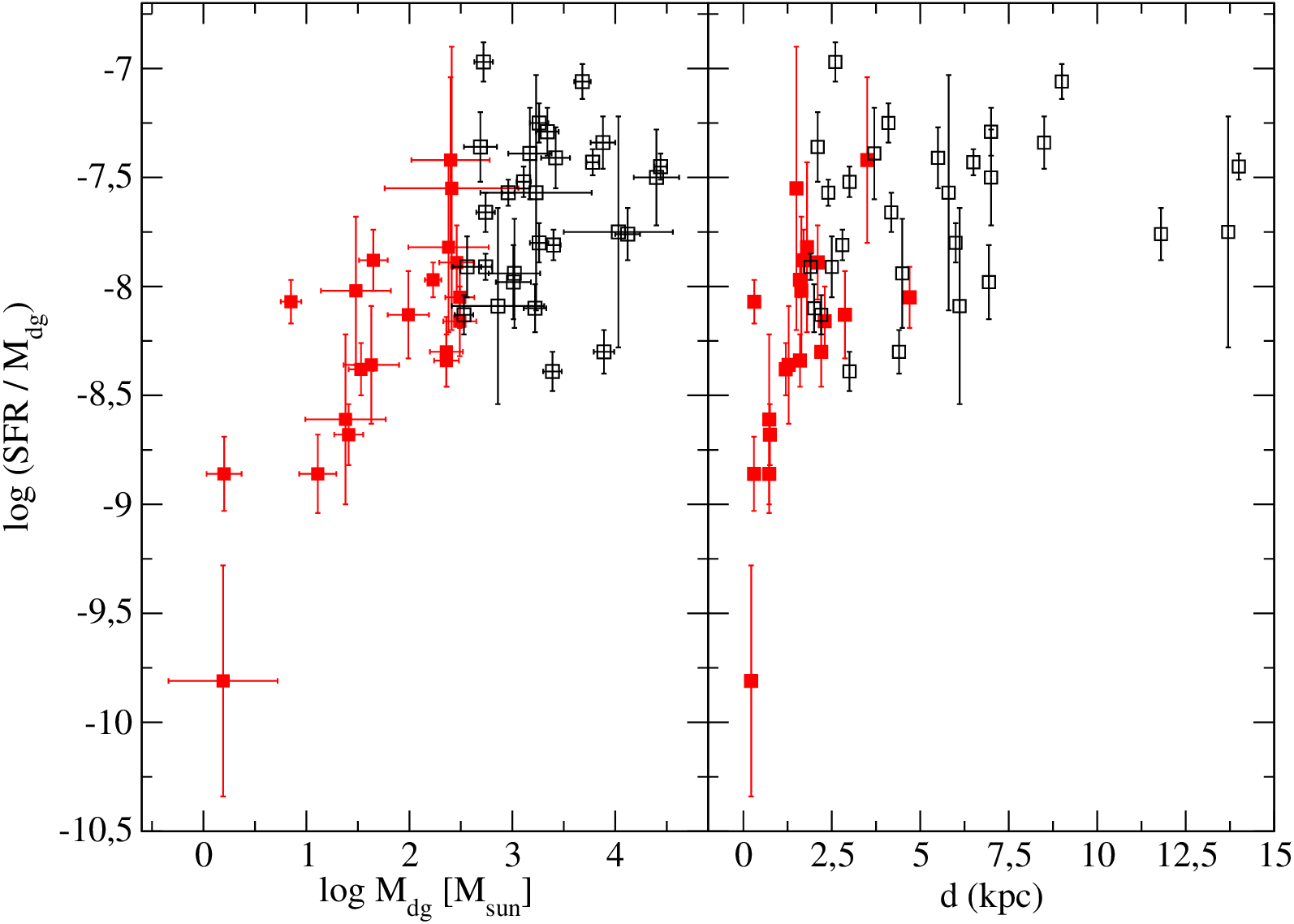}
\caption{SFR/M$_{dg}$ ratio versus the dense gas mass (left) and the distance (right) for all distant Galactic clouds from \citet{Wu10}. Red-filled and open squares correspond to log M$_{dg}$ $<$ and $>$ 2.50, respectively (log L$_{HCN}$ $<$ and $>$ 1.50).}
\label{Fig:Wu_data}
\end{figure}

Figures \ref{Fig:stars-clouds} and \ref{Fig:correlation} do not include the distant Galactic clouds from \citet{Wu10} with log M$_{dg}$ $<$ 2.50 (log L$_{HCN}$ $<$ 1.50). As shown in \citet{Wu05}, these clouds do not follow the linear correlation but have a super-linear slope. In other words, for a given M$_{dg}$ below that limit, distant Galactic clouds have a smaller SFR than predicted from the trend followed by the remaining Galactic clouds and the galaxies. Figure \ref{Fig:Wu_data} (left) shows that the ratio SFR/M$_{dg}$ of the clouds from \citet{Wu10} actually increases with M$_{dg}$ up to the mentioned limit, remaining roughly constant from that point. Based on this change of slope, \citet{Wu05} proposed the existence of a basic unit in star formation: if M$_{dg}$ is below the mass of that unit, SFR/M$_{dg}$ rises with M$_{dg}$ because clouds with low M$_{dg}$ do not sample the IMF completely by excluding high-mass stars. If M$_{dg}$ is above the mass of that unit, then the IMF is completely sampled and additional increases of M$_{dg}$ produce more units but not further increases in SFR/M$_{dg}$. In short, \citet{Wu05} proposed that the star formation correlation in other galaxies can be understood from the sum of basic units within each galaxy, these units being clouds with high-mass young stars. However, Fig. \ref{Fig:Wu_data} (right) shows that the ratio SFR/M$_{dg}$ also follows a trend with the distance to the clouds that is very similar to that with M$_{dg}$. The SFR/M$_{dg}$ ratio also increases for d $<$ 2.5 kpc, remaining roughly constant from that point. The Spearman's probability of false correlation accounting for the distances from the partial correlation technique \citep[e.g.,][]{Wall03} is 0.1 when only the 22 clouds with log M$_{dg}$ $<$ 2.50 from \citet{Wu10} are plotted in the SFR-M$_{dg}$ plane. This high p-value ($>$ 0.05) indicates that the SFR-M$_{dg}$ correlation for those sources is spurious since this is mainly driven by the d$^2$ factor used to compute both parameters compared. Therefore, it cannot be concluded that the change of the SFR/M$_{dg}$ ratio is caused by a physical limit in M$_{dg}$ because the influence of the distance cannot be ruled out. In conclusion, the proposal of high-mass star-forming clouds as the basic units driving the linear correlation cannot be supported from that data. 

In addition, the break of the SFR-M$_{dg}$ linear relation for clouds with low M$_{dg}$ could be explained from the fact that L$_{IR}$ was used as a tracer to measure the SFR. Most of the IR is emitted by high-mass stars and as soon as they are missing L$_{IR}$ could no longer be a good proxy of the SFR. In either case, by using different tracers for M$_{dg}$ and SFR \citet{Lada10,Lada12} already showed that the linear correlation is also valid for nearby clouds that do not necessarily host massive stars.

\section{Is the global correlation just the result of a large-scale, log-log diagram? }
\label{Sect:appendixc}
One may wonder if the enormous extent covered by the axes in Fig. \ref{Fig:correlation} somewhat hides the scatter and washes out significant trends within the data or, in other words, if any parameter could appear correlated simply because we are zooming out. Whereas the scale considered certainly has an effect on whether this or any other empirical trend is identified or not, not all data fit within the global correlation analyzed in this work. 

\begin{figure}
%[!hbtp]
\centering
 \includegraphics[width=8.5cm,clip=true]{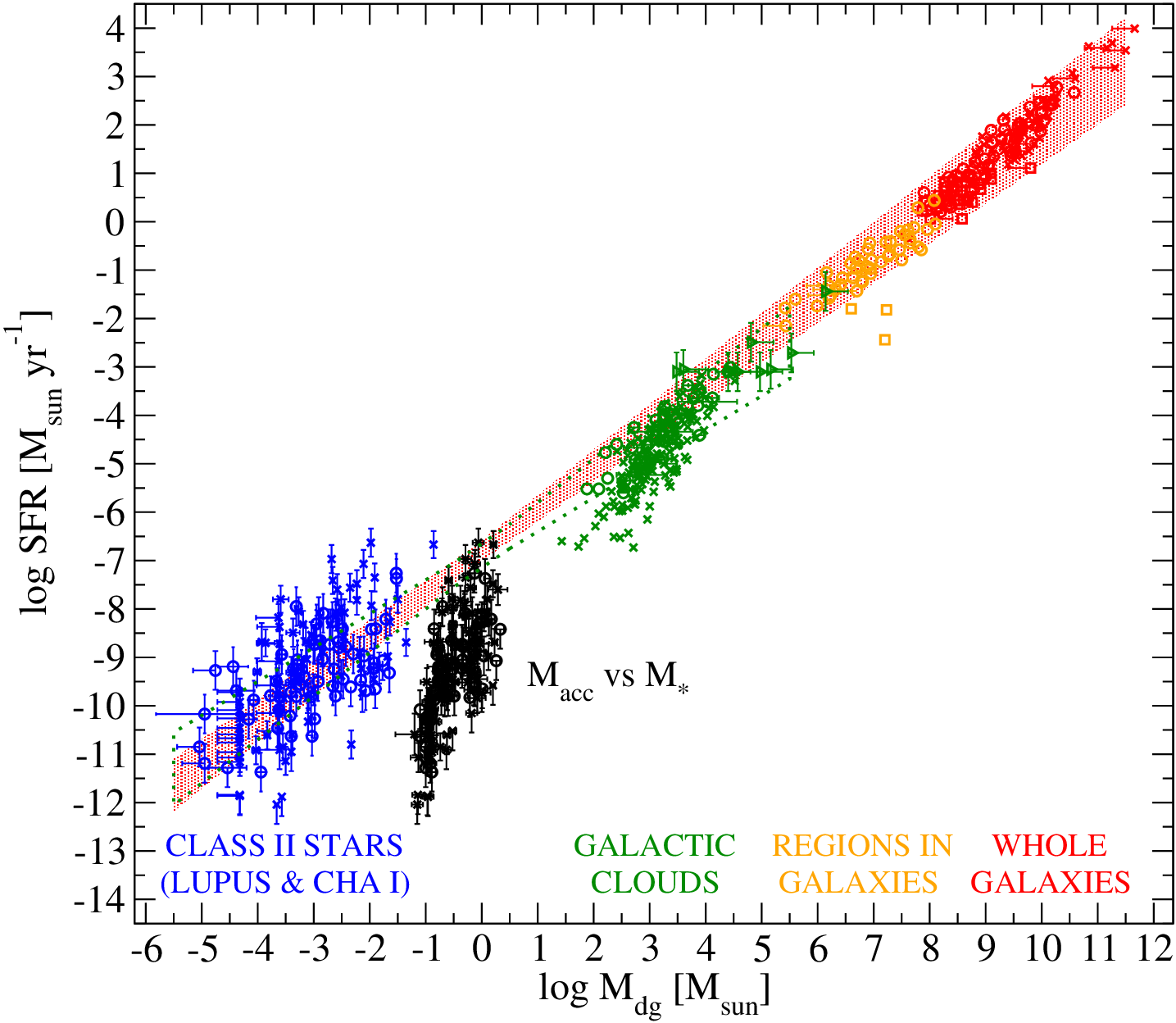}
\caption{Data and linear fits shown in Fig. \ref{Fig:correlation} including additional samples: Class II stars in the Cha I star-forming region \citep[blue crosses;][]{Mulders17}; massive Galactic clumps from the MALT90 survey \citep[green crosses;][]{Stephens16}; molecular clouds close to the Galactic center \citep[green triangles;][]{Kauffmann17a}; the central molecular zone (CMZ) in our Galaxy \citep[orange squares;][]{Longmore13}, regions within disks of other galaxies \citep[orange circles;][]{Usero15}, (U)LIRGS and star-forming galaxies \citep[red crosses and red squares;][]{GarciaBurillo12,Usero15}. On the other hand, the accretion rates in Lupus and Cha I are plotted versus the stellar masses instead of the disk masses \citep[black circles and crosses;][]{Manara16,Mulders17}.}
\label{Fig:correlation_extended}
\end{figure}

We have compiled additional data on top of that analyzed in Fig. \ref{Fig:correlation}, which is plotted in Fig. \ref{Fig:correlation_extended} (references included in the caption). First, recent ALMA measurements of Class II stars in the Cha I star-forming region (blue crosses) have been added to the Lupus sources, both constituting the only complete data for Class II stars available to date. Even when the origin of the large scatter in $\dot{M}_{\rm acc}$ is subject to debate \citep[accretion variability, observational limits, sample selection effects, different evolutionary stages and initial conditions in previous phases; see, e.g.,][and references therein]{Mendi15,Manara16,Mulders17}, it has been shown that there is a statistically significant, roughly linear correlation in each star-forming region separately \citep{Manara16,Mulders17}. Indeed, considering the Lupus and Cha I samples together, we derive a slope $\sim$ 0.9 $\pm$ 0.2. Concerning the rest of the new data in Fig. \ref{Fig:correlation_extended}, the samples are selected because they have been reported to show deviations with respect to the linear scaling relation. For instance, the additional sample of (U)LIRGS (red crosses) shows a slightly super-linear slope. On the contrary, pointings of specific regions in disks of other galaxies \citep[orange circles, see also, e.g.,][]{Bigiel16} show a sub-linear slope, as well as clouds close to the Galactic center (green triangles with error bars) when these are considered alone. These sub-trends are relatively washed out under the axes range considered here and are roughly incorporated to the global correlation. In addition, the central molecular zone (orange squares) and some massive clumps in our Galaxy from the Millimetre Astronomy Legacy Team 90 GHz (MALT90) survey (green crosses with $\sim$ log M$_{dg}$ $<$ 4) appear displaced from the global correlation even when such a large axes range is used. Despite the fact that the tracers and methodology used to derive the SFR and M$_{dg}$ values in many of the samples included in Fig. \ref{Fig:correlation_extended} are the same as those described in Appendix \ref{Sect:appendixa}, different approaches have also been used \citep[see, e.g.,][]{Longmore13,Kauffmann17a}. The recent work by \citet{Shimajiri17} may solve most apparent deviations with respect to the global correlation from a careful determination of the conversion factors between the observables and the physical parameters involved \citep[see also, e.g.,][]{Lada12,Tan18}, which have not been considered in this work. However, even if the correlation changes in certain environments, the scope of this paper is not to explain the possible sub-trends within the global correlation but to point out that the individual stars could eventually play a role in such an explanation by means of the bottom-up hypothesis.  

In contrast, Fig. \ref{Fig:correlation_extended} also shows that when the stellar masses are plotted instead of the disk masses (black circles and crosses), the well known $\dot{M}_{\rm acc}$-M$_*$ correlation \citep[e.g.,][and references therein]{Mendi15,Mulders17} shows up with a very different slope and intercept, and can be clearly distinguished from the global correlation linking stars, clouds, and galaxies. This simple modification in one of the axes demonstrates that not any single stellar parameter appears to be correlated in a logarithmic scale plot spanning over a huge axes range.

\section{Relation between least-squares intercepts for stars, clouds, and galaxies}
\label{Sect:appendixd}
The $\dot{M}_{\rm acc}$-M$_{disk}$ linear relation observed in Class II stars can be generalized to also include the total circumstellar gas mass in Class 0/I sources. For a sample of young stars in a given cloud, this is expressed as log $\dot{M}_{\rm acc}$ = log K$_s$ + log M$_{cs}$. If least-squares linear regression is used, by definition the intercept is given by log K$_s$ = $<$log $\dot{M}_{\rm acc}$$>$ - $<$log M$_{cs}$$>$, where the arithmetic means refer to the total number of stars in the cloud (N$_s$). The value of K$_s$ can then be expressed as
\begin{equation}
\label{Eq.stars}
K_s = \left(\prod_{i=1}^{N_s}\frac{\dot{M}_{\rm acc_i}}{M_{cs_i}}\right)^{1/N_s} = \left(\prod_{i=1}^{N_s} k_{s_i}\right)^{1/N_s},
\end{equation}
with k$_{s}$$_{i}$ = $\dot{M}_{\rm acc}$$_i$/M$_{cs}$$_i$ the ratio between the accretion rate and the circumstellar gas mass for each star. Therefore, K$_s$ is the geometric mean of the individual k$_s$$_i$ ratios for each star. Similarly, the least-squares intercept derived for a sample of N$_c$ clouds provides a K$_c$ value that is the geometric mean of the individual SFR$_j$/M$_{dg}$$_j$ ratios of each cloud, and equivalently for a sample of galaxies. Under the hypothesis proposed in this work, the SFR of a given cloud is given by the sum of the individual accretion rates of the stars within that cloud. Similarly, the M$_{dg}$ values are given by the sum of the total circumstellar gas masses of the stars within that cloud, divided by a factor $f$ that accounts for the fraction of dense gas mass that is directly involved in star formation (i.e., for a given cloud $\Sigma$M$_{cs}$ = $f$M$_{dg}$). Therefore,
\begin{equation}
\label{Eq.clouds}
K_c = \left(\prod_{j=1}^{N_c}\frac{f_j<\dot{M}_{\rm acc}>_j}{<M_{cs}>_j}\right)^{1/N_c} = \left(\prod_{j=1}^{N_c} f_jk_{c_j}\right)^{1/N_c}.
\end{equation}
Similarly, for a sample of N$_g$ galaxies,
\begin{equation}
\label{Eq.galaxies}
K_g = \left(\prod_{k=1}^{N_g}\frac{f_k<\dot{M}_{\rm acc}>_k}{<M_{cs}>_k}\right)^{1/N_g} = \left(\prod_{k=1}^{N_g} f_kk_{g_k}\right)^{1/N_g}.
\end{equation}
Equations \ref{Eq.stars}, \ref{Eq.clouds}, and \ref{Eq.galaxies} show that the intercepts of clouds and galaxies are determined mainly by the ratio between the accretion rate and the circumstellar gas mass of the young stars hosted by them. The simplest scenario occurs when this ratio is the same for all stars, that is, k$_s$$_i$ = k$_s$ = $\dot{M}_{\rm acc}$/M$_{cs}$, and the fraction of dense gas directly participating in stellar accretion is the same for all clouds and galaxies. In this case $f$K$_s$ = K$_c$ = K$_g$, that is, the intercepts of clouds and galaxies are the same as for the stars only if the amount of gas directly participating in star formation is 100$\%$ of the dense gas in clouds and galaxies ($f$ = 1), and smaller than the stars otherwise (0 $<$ $f$ $<$ 1). From a physical perspective, $\tau$$_s$ = k$_s$$^{-1}$ is the accretion timescale on which all M$_{cs}$ would fall onto a given star for a constant mass accretion rate $\dot{M}_{\rm acc}$. Under the assumption of equal accretion to circumstellar gas ratios and $f$-factors, $\tau$$_s$/$f$ = $\tau$$_c$ = $\tau$$_g$, which means the gas depletion timescales for clouds and galaxies are the same as the accretion timescales for the stars only if the amount of gas directly participating in star formation is 100$\%$ of the dense gas in clouds and galaxies ($f$ = 1), and larger than the stars otherwise (0 $<$ $f$ $<$ 1). 

Finally, when a single cloud is considered, Eq. \ref{Eq.clouds} reduces to K$_c$ = $f$k$_c$ = $f$$<$$\dot{M}_{\rm acc}$$>$/$<$M$_{cs}$$>$ = $\sum$$_{i=1}^{N_s}$ $f$k$_s$$_i$M$_{cs}$$_i$/$\sum$$_{i=1}^{N_s}$ M$_{cs}$$_i$. For a given M$_{cs}$ value, the scatter in $\dot{M}_{\rm acc}$ for the stars can be quantified through a fixed value $\sigma$k$_s$$_i$ = $\sigma$k$_s$, which translates to 
\begin{equation}
\label{Eq.scatter}
\sigma k_c = \sigma k_s \times \left(\frac{\sum_{i=1}^{N_s} M_{cs_i}^2}{\left(\sum_{i=1}^{N_s} M_{cs_i}\right)^2}\right)^{1/2} < \sigma k_s.
\end{equation}
Consequently, for the galaxies, $\sigma$k${_g}$ $<$ $\sigma$k${_c}$. Therefore, the y-axis scatter in the global correlation should be maximum for stars and decreases from clouds to galaxies.

\end{appendix}

\begin{thebibliography}{}
\bibitem[Audard et al.(2014)]{Audard14} Audard, M.; Ábrah\'am, P.; Dunham, M.M. et al. 2014, PPVI, 387-410
\bibitem[Benedettini et al.(2012)]{Benedettini12} Benedettini, M.; Pezzuto, S.; Burton, M.G. et al. 2012, MNRAS, 419, 238
\bibitem[Bergin et al.(2007)]{Bergin07} Bergin, E.A.; Aikawa, Y.; Blake, G.A.; van Dishoeck, E.F. 2007, PPV, 751-766
\bibitem[Bigiel et al.(2016)]{Bigiel16} Bigiel, F.; Leroy, A.K.; Jim\'enez-Donaire, M.J. et al. 2016, ApJ, 822, L26
\bibitem[Bontemps et al.(1996)]{Bontemps96} Bontemps, S.; Andr\'e, Ph.; Terebey, S.; Cabrit, S. 1996, A\&A, 311, 858
\bibitem[Carney et al.(2016)]{Carney16} Carney, M.T.; Yildiz, U.A.; Mottram, J.C. et al. 2016, A\&A, 586, A44
\bibitem[Chomiuk \& Povich(2011)]{Chomiuk11} Chomiuk, L.; Povich, M.S. 2011, AJ, 142, 197
\bibitem[Dullemond et al.(2006)]{Dullemond06} Dullemond, C.P,; Natta, A.; Testi, L. 2006, ApJ, 645, L69
\bibitem[Enoch et al.(2008)]{Enoch08} Enoch, M.L.; Evans II, N.J.; Sargent, A.I. et al. 2008, ApJ, 684, 1240
\bibitem[Fedele et al.(2010)]{Fedele10} Fedele, D.; van den Ancker, M.; Henning, Th.; Jayawardhana, R.; Oliveira, J.M. 2010, A\&A, 510, A72
\bibitem[Gao \& Solomon(2004)]{Gao04} Gao, Y. \& Solomon, P.M. 2004, ApJ, 606, 271
\bibitem[Garc\'\i{}a-Burillo et al.(2012)]{GarciaBurillo12} Garc\'\i{}a-Burillo, S.; Usero, A.; Alonso-Herrero, A. et al. 2012, A\&A, 539, A8 
\bibitem[Herczeg et al.(2017)]{Herczeg17} Herczeg, G.J.; Johnstone, D.; Mairs, S. et al. 2017, ApJ, 849, 43
\bibitem[Jackson et al.(2013)]{Jackson13} Jackson, J.M.; Rathborne, J.M.; Foster, J.B. et al. 2013, PASP, 30, 57
\bibitem[Kauffmann et al.(2017b)]{Kauffmann17b} Kauffmann, J.; Goldsmith, P.F.; Melnick, G. et al. 2017b, A\&A, 605, L5
\bibitem[Kauffmann et al.(2017a)]{Kauffmann17a} Kauffmann, J.;Pillai, T.; Zhang, Q. et al. 2017a, A\&A, 603, A89
\bibitem[Kennicutt \& Evans(2012)]{Kennicut12} Kennicutt, R.C. \& Evans, N.J. 2012, ARA\&A, 50, 531
\bibitem[Kennicutt(1998)]{Kennicutt98} Kennicutt, R.C. 1998, ApJ, 498, 541
\bibitem[K\"onyves et al.(2015)]{Konyves15} K\"onyves, V.; Andr\'e, Ph.; Men'shchikov, A. et al. 2015, A\&A, 584, A91
\bibitem[Lada(1987)]{Lada87}Lada, C.J. 1987, Star Forming Regions, 115, 1
\bibitem[Lada et al.(2010)]{Lada10} Lada, C.~J; Lombardi, M.; Alves, J.F. 2010, ApJ, 724, 687
\bibitem[Lada et al.(2012)]{Lada12} Lada, C.~J; Forbrich, J.; Lombardi, M.; Alves, J.F. 2012, ApJ, 745, 190
\bibitem[Lada et al.(2017)]{Lada17} Lada, C.~J.; Lewis, J.A.; Lombardi, M.; Alves, J. 2017, A\&A, 606, 100
\bibitem[Lewis \& Lada(2016)]{Lewis16} Lewis, J.A. \& Lada, C.~J. 2016, ApJ, 825, 91
\bibitem[Longmore et al.(2013)]{Longmore13} Longmore, S.N.; Bally, J.; Testi, L. et al. 2013, MNRAS, 429, 987
\bibitem[Lombardi et al.(2014)]{Lombardi14} Lombardi, M.; Bouy, H. ; Alves, J.; Lada, C.J. 2014, A\&A, 566, A45 
\bibitem[Manara et al.(2016)]{Manara16} Manara, C.F.; Rosotti, G.; Testi, L. et al. 2016, A\&A, 591, L3
\bibitem[Megeath et al.(2012)]{Megeath12} Megeath, S.T.; Gutermuth, R.; Muzerolle, J. et al. 2012, AJ, 144, 192
\bibitem[Mendigut\'\i{}a et al.(2012)]{Mendi12} Mendigut\'\i{}a, I.; Mora, A.; Montesinos, B. et al. 2012, A\&A, 543, A59 
\bibitem[Mendigut\'\i{}a et al.(2015)]{Mendi15} Mendigut\'\i{}a, I.; Oudmaijer, R.D.; Rigliaco, E. et al. 2015, MNRAS, 452, 2837
\bibitem[Mer\'\i{}n et al.(2008)]{Merin08} Mer\'\i{}n, B.; J\o{}rgensen, J.; Spezzi, L. et al. 2008, ApJS, 177, 551
\bibitem[Mowat et al.(2017)]{Mowat17} Mowat, C.; Hatchell, J.; Rumble, D. et al. 2017, MNRAS, 467, 812
\bibitem[Mulders et al.(2017)]{Mulders17} Mulders, G.D.; Pascucci, I.; Manara, C.F. et al. 2017, ApJ, 847, 31
\bibitem[Najita et al.(2007)]{Najita07} Najita, J.R.; Strom, S.E.; Muzerolle, J. 2007, MNRAS, 378, 369
\bibitem[Najita et al.(2015)]{Najita15} Najita, J.R.; Andrews, S.M.; Muzerolle, J. 2015, MNRAS, 450, 3559
\bibitem[Padoan et al.(2014)]{Padoan14} Padoan, P.; Haugb\o{}lle, T.; Nordlund, A. 2014, ApJ, 797, 32
\bibitem[Palla \& Stahler(2000)]{Palla00} Palla, F. \& Stahler, S.W. 2000, ApJ, 540, 255
\bibitem[Pety et al.(2017)]{Pety07} Pety, J.; Guzm\'an, V.V.; Orkisz, J.H. et al. 2017, A\&A, 599, A98
\bibitem[Pomohaci et al.(2017)]{Pomohaci17} Pomohaci, R.; Oudmaijer, R.D.; Lumsden, S.L.; Hoare, M.G.; Mendigut\'\i{}a, I. 2017, MNRAS, 472, 3624
\bibitem[Rigliaco et al.(2015)]{Rigliaco15} Rigliaco, E.; Pascucci, I.; Duchene, G. et al. 2015, ApJ, 801, 31
\bibitem[Rosotti et al.(2017)]{Rosotti17} Rosotti, G.P.; Clarke, C.J.; Manara, C.F.; Facchini, S. 2017, MNRAS, 468, 1631
\bibitem[Rygl et al.(2013)]{Rygl13} Rygl, K.L.J.; Benedettini, M.; Schisano, E. et al. 2013, A\&A, 549, L1
\bibitem[Schmidt(1959)]{Schmidt59} Schmidt, M. 1959, ApJ, 129, 243
\bibitem[Shimajiri et al.(2017)]{Shimajiri17} Shimajiri, Y.; Andr\'e, Ph.; Braine, J. et al. 2017, A\&A, 604, A74
\bibitem[Stephens et al.(2016)]{Stephens16} Stephens, I.W.; Jackson, J.M.; Whitaker, J.S. et al. 2016, ApJ, 824, 29
\bibitem[Tafalla et al.(1998)]{Tafalla98} Tafalla, M.; Mardones, D.; Myers, P.C. et al. 1998, ApJ, 504, 900
\bibitem[Tan et al.(2018)]{Tan18} Tan, Q.-H.; Gao, Y.; Zhang, Z.-Y. et al. 2018, ApJ, 860, 165
\bibitem[Usero et al.(2015)]{Usero15} Usero, A.; Leroy, A.K.; Walter, F. et al. 2015, AJ, 150, 115 
\bibitem[Vorobyov \& Basu(2008)]{Vorobyov08} Vorobyov, E.I. \& Basu, S. 2008, ApJL, 676, L139 
\bibitem[Wall \& Jenkins(2003)]{Wall03} Wall, J.V. \& Jenkins, C.R. 2003, Practical statistics for astronomers
(Cambridge University Press)
\bibitem[Wu et al.(2005)]{Wu05} Wu, J.; Evans, N.J., II; Gao, Y. et al. 2005, ApJ, 635, L173
\bibitem[Wu et al.(2010)]{Wu10} Wu, J.; Evans, N.J., II; Shirley, Y. L.; Knez, C. 2010, ApJS, 188, 313
\bibitem[Zari et al.(2016)]{Zari16} Zari, E.; Lombardi, M.; Alves, J.; Lada, C.J.; Bouy, H. 2016, A\&A, 587, A106

\end{thebibliography}
\end{document}